\documentclass{bmcart}

\usepackage[utf8]{inputenc} 
\usepackage{makecell}
\usepackage{mdwlist}
\usepackage{amsmath}
\usepackage{amsfonts}
\usepackage{graphicx}
\usepackage{enumerate}

\startlocaldefs%

\newcommand{\Pb}{\mathbb{P}}

\newcommand{\btheta}{\mathbf{\theta}}
\newcommand{\bX}{\mathbf{X}}
\newcommand{\bZ}{\mathbf{Z}}

\endlocaldefs

\begin{document}

\begin{frontmatter}

\begin{fmbox}
\dochead{Methodology}


\title{A model for gene deregulation detection using expression data}


\author[
   addressref={aff1},                   
   email={thomas.picchetti@parisdescartes.fr}   
]{\inits{T}\fnm{Thomas} \snm{Picchetti}}
\author[
   addressref={aff2},
   email={julien.chiquet@genopole.cnrs.fr}
]{\inits{J}\fnm{Julien} \snm{Chiquet}}
\author[
   addressref={aff3},
   email={Mohamed.Elati@issb.genopole.fr}
]{\inits{M}\fnm{Mohamed} \snm{Elati}}
\author[
   addressref={aff2},
   email={pierre.neuvial@genopole.cnrs.fr}
]{\inits{P}\fnm{Pierre} \snm{Neuvial}}
\author[
   addressref={aff3,aff4},
   email={remy.nicolle@issb.genopole.fr}
]{\inits{R}\fnm{R\'emy} \snm{Nicolle}}
\author[
   addressref={aff1},
      corref={aff1},                       
   email={etienne.birmele@parisdescartes.fr}
]{\inits{E}\fnm{Etienne} \snm{Birmel\'e}}


\address[id=aff1]{
  \orgname{Laboratoire MAP5, Universit\'e Paris Descartes and CNRS, Sorbonne Paris Cit\'e}, 
  \street{45 rue des Saints-P\`eres},                     %
  \postcode{75270}                                
  \city{Paris Cedex 06},                              
  \cny{France}                                    
}
\address[id=aff2]{%
  \orgname{Laboratoire de Mathématiques et Modélisation d’Evry (LaMME)},
  \street{Université d’Evry-Val-d’Essonne/UMR CNRS 8071/ENSIIE/USC INRA},
  \postcode{}
  \city{Evry},
  \cny{France}
}
\address[id=aff3]{%
  \orgname{institute of Systems and Synthetic Biology (iSSB)},
  \street{CNRS},
  \postcode{}
  \city{University of Evry},
  \cny{France}
}
\address[id=aff4]{%
  \orgname{Institut Curie, PSL Research University},
  \street{UMR 144 75248 Cedex 05, France},
  \postcode{CNRS}
  \city{75248 Paris Cedex 05},
  \cny{France}
}


\begin{artnotes}
\end{artnotes}

\end{fmbox}


\begin{abstractbox}

\begin{abstract} 
  In tumoral  cells, gene regulation mechanisms  are severely altered.
Genes that do not react normally to their regulators' activity can
  provide explanations for the tumoral behavior, and be characteristic
  of cancer subtypes. We thus propose a  statistical  methodology  to
  identify the misregulated  genes given a reference  network and gene
  expression data.

  Our model  is based on a  regulatory process in which  all genes are
  allowed to  be deregulated. We  derive an EM algorithm  where the
  hidden  variables  correspond  to  the  status  (under/over/normally
  expressed) of the  genes and where the E-step is  solved thanks to a
  message  passing   algorithm.   Our  procedure   provides  posterior
  probabilities  of deregulation  in a  given sample  for each
  gene.   We  assess  the  performance  of  our  method  by  numerical
  experiments on simulations and on a bladder cancer data set.

\end{abstract}


\begin{keyword}
\kwd{regulatory network}
\kwd{belief propagation}
\kwd{EM algorithm}
\kwd{deregulation}
\kwd{inference}
\end{keyword}


\end{abstractbox}
%

\end{frontmatter}



\section*{Background}
Various mechanisms affect gene expression in tumoral cells, including
copy number alterations, mutations, modifications in the regulation
network between the genes. A simple strategy to identify genes affected
by these phenomena is to perform differential expression analysis. Results can
then be extended to the scale of pathways using enrichment analysis
\cite{KDO03} or functional class scoring \cite{STM05}. However,  such
a  strategy  is  blind to  small  variations  in  gene expression,
especially   as  multiple  testing   correction  applies.  Moreover,
it does not take interdependence between genes into account
and can mark an expression change as abnormal when
actually it is induced by a change in the regulators' activity.
To overcome  these drawbacks,  an alternative
strategy is to identify the affected genes by pointing important
changes in the gene regulatory network (GRN) of the tumoral cell.
Such an approach furthermore corresponds to the modelisation of
phenomena altering regulation, as for instance mutations in regulatory regions \cite{MRS15}.

The first step towards this is to procure a GRN.
It can  be obtained  from curated  databases or,  in order  to obtain
tissue or  condition-specific networks, reconstructed  from expression
data. In the latter case, the  inference can be done by relying either
on discrete  or continuous  models.  In  the discrete  framework, gene
expression profiles are discretized into binary or ternary valued variables
(underexpressed/normal/overexpressed).    The   regulation
structure is  then given by  a list of truth  tables~\cite{ER11}. This
approach allows in particular to  take coregulation into account, that
is  to  require the  activity  of  a  whole  set of  co-activators  or
co-inhibitors to  activate or  inhibit the  target \cite{ENB07,NRE15}.
In  the  continuous  case,  inference  can be  done  in  a  regression
framework, where  the expression of  each target gene is  explained by
all its  potential regulator  genes.  An edge  is drawn  between two
genes  if the  corresponding regression  coefficient is  significantly
different from  zero, which can  be deciphered by  performing variable
selection in the  regression model. A popular choice for  this task is
to  rely  on   sparsity-inducing  penalties  like  the   Lasso  and  its
by-products \cite{HMV12,MB06}.  In particular,  some variants allow to
account for co-regulation by  favoring predefined groups of regulators
acting together in  a sign-coherent way \cite{CGC12}.   Other forms of
penalties    encourage   a    predefined    hierarchy   between    the
predictors \cite{JAB11},  \emph{i.e.}  the regulator  genes in  the
case at hand.

To unravel deregulated  genes by means of GRN, a  first possibility is
to infer several  networks independently (one for each  tissue) and to
compare them.  However, due to the noisy nature of transcriptomic data
and the large number of features  compared to the sample size, most of
the differences found  in the networks inferred  independently may not
be linked with underlying  biological processes.  Methods have therefore
been developed to infer several networks jointly to share similarities
between the different tissues and penalize  the presence of an edge in
only  one  of   them.   Such  methods  exist  for   both  time  series
\cite{KIY12} or steady-state \cite{CGA11} data.

A second possibility  is to assess the adequacy of  gene expression in
tumoral cell to a reference  GRN, in order to exhibit the more
striking discrepancies  -- \emph{i.e.}  the regulations  which are not
fulfilled by  the data  --.  In this  perspective, \cite{KS12}  use an
heuristic in a Boolean framework to update the regulatory structure by
minimizing the discrepancies between the  reference GRN and a new data
set.  A  similar approach is  depicted in \cite{GBM09} to  predict the
discrepancies and the  unobserved genes of the  network.  More methods
analyzing the coherence between known signaling pathways and gene data
sets can  be found in  the review  \cite{SK13}.  Still, they  focus on
checking the validity  of the network rather  than highlighting genes
with an abnormal behavior.

At the pathway level rather than the gene level, it is possible to look
for   sample-specific   regulation   abnormalities   by   using   SPIA
\cite{TDP09}.    PARADIGM    \cite{VSB10}   generalizes   SPIA   on
heterogeneous data (DNA  copies, mRNA and protein  data). Moreover, it
determines a  score of activity  for each gene  of a pathway  for each
sample of  the data  set, and  the use of  hidden variables  allows to
compute this score  even if some of  the genes of the  pathway are not
measured. The  method is  however not network-wide  in the  sense that
each gene has a deregulation score by pathway it belongs to, and
pathways  are  treated independently.  Moreover,  as  the pathways  are
extracted from  curated databases,  the regulations taken  into account
are not tissue-specific.
\\*

The aim of this paper is to  develop a methodology to provide a network-wide
deregulation  score  for each  gene and each sample by  taking the  whole  regulation
network into account. For this purpose,  we introduce a model based on
a regulatory  process in  which genes are  allowed to  be deregulated,
\emph{i.e.}   not respond  to  their regulators  as  expected.  An  EM
strategy  is  proposed  for  parameter  inference,  where  the  hidden
variables correspond to the  status (under/over/normally expressed) of
the  genes.   The  E-step  is  solved  thanks  to  a  message  passing
algorithm.   At   the  end   of  the   day,  the   procedure  provides
\emph{posterior} probabilities  of deregulation in a  given sample for
each  target  gene.  We  assess  the  performance  of our  method  for
detecting  deregulations on  simulated  data. We  also illustrate  its
interest  on   a  bladder  cancer   data  set,  where  we   study  the
deregulations  according   to  two  reference  GRN   obtained  by  two
state-of-the-art   network  inference   procedures   on  a   consensus
expression data set.

\section*{Methods}

\subsection*{The model}

Our  model  draws  inspiration   from  LICORN  \cite{ENB07},  a  model
originally developed for network  inference purposes. LICORN considers
a  regulation   structure  in   which  genes  are   either  regulators
(transcription factors --  TFs) or target genes.   The expressions are
discretized and each gene $g$ is characterized by a ternary value $S_g
\in \{-1,0,+1\}$  encoding its expression status  -- under-, normally,
or over-expressed. The regulation of  each target gene $g$ is governed
by a  set of co-activators  $A(g)$ and co-inhibitors $I(g)$  among the
TFs. Those sets are endowed  with some ``collective status'' described
by variables $S_g^A$ and $S_g^I$,  assuming that regulation works in a
cooperative way: hence, the collective state of a set of regulators is
over- (resp. under-)  represented if and only if all  elements in the
set share  the same status.  Finally,  the status $S_g$ of  the target
gene $g$  is deduced from $S_g^A$  and $S_g^I$ by following Truth
Table~\ref{licorntable}.

In order to detect deregulated target genes given a regulatory network
and gene expression  profiles, we apply two major modifications to the
LICORN  model:  first,   we  avoid  discretization  of   the  data  by
considering  all the  ternary variables  introduced so  far as  hidden
random variables.   The expression $X_g$ of  a gene $g$ is  assumed to
follow a normal distribution with parameters that depend on the hidden
status,  \emph{i.e.},  $X_g  \vert   S_g=s  \sim  \mathcal{N}(  \mu_s,
\sigma_s)$.  Second, we introduce for  each gene an indicator variable
$D_g$   for  deregulation,   such   that   $D_g=1$  with   probability
$\epsilon$. Renaming  the result  of the truth  table by  $S_g^R$, the
final status  of the target is  then deduced from the  values of $D_g$
and $S_g^R$:
\begin{displaymath}
\begin{cases}
      S_g=S_g^R & \textrm{if } D_g=0, \\
      \forall s\neq S_g^R, \Pb(S_g=s) = \frac{1}{2} & \textrm{if } D_g=1.
    \end{cases}
\end{displaymath}

For  completeness, we  must  specify the  distribution  of the  hidden
states  $S_g$   for  each  TFs:  we   assume  independent  multinomial
distributions                      with                     parameters
$\boldsymbol\alpha=(\alpha_-,\alpha_0,\alpha_+)$.

The model  is summarized for one target gene in Figure \ref{modelfig}.
For the  sake of
conciseness,  the  vector  $\boldsymbol\theta$  entails all  parameters  of  the
models, that is,  the means and standard deviations  of the Gaussians,
the  vector  $\boldsymbol\alpha$  of   proportions  and  the  deregulation  rate
$\epsilon$.   The data set  contains  $n$ samples,  $r$ TFs  and $t$
target genes.  We denote by $\bZ$  the $n \times (r+5t)$ matrix of all
hidden  states  and  by  $\bX$  the  $n\times  (r+t)$  matrix  of  all
expression variables.

Note that the dependencies among variables are acyclic, implying that the likelihood can be decomposed in a product.
\begin{multline*}
p(X,Z|\theta)=\prod p(S_j | \alpha)  \times \prod p(S^A_i|S_j\ldots)
\times \prod p(S^I_i|S_j\ldots) \times \prod p(S^R_i |S^I_i, S^A_i) \\
\times \prod p(D_i|\epsilon) \times \prod p(S_i|S^R_i,D_i) \times \prod p(X_k|S_k,\mu,\sigma)
\end{multline*}
For sake of readability, the indices of the products are omitted in the above formula. However, it should be clear when the product runs over target genes, regulator genes or all of them.

\subsection*{Estimation algorithm}

As usual with latent variable models, the likelihood is intractable as
the  number  of  potential  states   of  the  hidden  variables  grows
exponentially with  the number of  variables.  Therefore, we  adopt an
EM-like strategy \cite{emref} by iterating the following steps,
starting from  an initial  guess $\boldsymbol\theta^0$  of the  model parameters:
\begin{description}
\item[E-step:] Fix $\boldsymbol\theta$ and compute the conditional probability distribution of
  the hidden variables, given the observed expression values:
  $q(\bZ)=\Pb(\bZ|\bX,\boldsymbol\theta)$
\item[M-step:] Fix $q$ and find $\boldsymbol\theta$ that maximizes $\sum
  q(\bZ) \log\Pb(\bX,\bZ\vert \boldsymbol\theta)$
\end{description}

\paragraph*{Step E.} The first issue at stake in the E-step is to deal
with the  number of potential states  for the hidden variables  of all
the genes.  Fortunately, we only  need their marginal distributions in
the M step, as will be shown in the corresponding section.
Still, we need a way to compute these marginals without having to
compute the joint distribution first.

To   handle  this  issue,   we  rely   on  Belief   Propagation
\cite{yedidia} -- a.k.a \emph{message-passing algorithm} -- to perform
the E step, since the  probability distribution arising from our model
is easily  represented as a factor  graph.  Indeed, consider a  set of
discrete values for all variables $S^A_g$, $S^I_g$, $S^R_g$ and $D_g$.
Conditionally on $\bX$, the probability  for the discrete variables to
match the given value is proportional to the product of the following
factors:
\begin{enumerate}[1.]
\item $\alpha_{S_g}$ for each regulator gene $g\in R$;
\item $\epsilon$ if $D_g=1$, and $\frac{1-\epsilon}{2}$ if $D_g=0$, for each target gene $g \in T$;
\item $\frac{1}{\sigma}\exp\frac{-(X_g-\mu)^2}{2\sigma^2}$ for each gene $g \in G$ (regulator or target), where $\mu$ and $\sigma$ are the mean expression and standard deviation associated to state $S_g$;
\item a factor equal to one if $S^A_g$ correctly represents the collective state of $g$'s activators, and zero otherwise;
\item a factor equal to one if $S^I_g$ correctly represents the collective state of $g$'s inhibitors, and zero otherwise;
\item a factor equal to one if $S^R_g$ is the entry in Table \ref{licorntable} corresponding to $S^A_g$ and $S^I_g$, and zero otherwise;
\item a factor equal to one if either $D_g=0$ and $S_g=S^R_g$ or $D_g=1$ and $S_g\not = S^R_g$, and zero otherwise.
\end{enumerate}

This  factorization  translates  into  the factor  graph  depicted  in
Figure~\ref{factorgraph} (a  graph whose  nodes are the  variables and
the above  factors, each  factor being connected  to the  variables it
depends  on).    We  use  the  \emph{SumProduct}   Belief  Propagation
algorithm, implemented in the  Dimple library \cite{dimple} to compute
approximated marginals of every hidden variable, given the regulation
network, the  parameter set,  and the expression  values. In  the case
where multiple  samples are given,  this can be done separately for
each one since the samples are considered as independent.

\paragraph*{Step M.} In this step we keep the probability distribution
$q$ fixed and look for the parameters $\boldsymbol\theta$ that maximize
\begin{displaymath}
\sum_\bZ q(\bZ) \log \Pb(\bX,\bZ\vert \btheta)
\end{displaymath}
Since $\Pb(\bX,\bZ\vert \btheta)$ is a  product of simple factors, its
logarithm is the sum of these factors. Also, note that boolean factors
(4-7) can  be omitted since they  have no effect on  the sum: whenever
$q(Z)\not =0$, these factors must be equal to 1 hence the logarithm is 0.

Calling $G$ the  set of genes, $R\subset G$ the  set of regulators and
$T\subset G$ the set of target genes,  we are left to maximize the sum
over all samples of
\begin{multline*}
\sum_{g\in R} \sum_Z q(Z)\log\alpha_{S_g}\\
 + \sum_{g\in T} \sum_Z q(Z)\left(D_g\log\epsilon+(1-D_g)\log\frac{1-\epsilon}{2}\right)\\
+ \sum_{g\in G} \sum_Z q(Z)\left(\frac{-(X_g-\mu_{S_g})^2}{2\sigma_{S_g}^2}-\log\sigma_{S_g}\right)
\end{multline*}

These three terms  depend on separate parameters and  can be maximized
separately. Moreover, we only require the marginals of variables $S_g$
and $D_g$ for  this task, and not the full  distribution $q$. Denoting
by $I$ the set of samples, it is straightforward to show that the former sum is maximized for the following parameters:
\begin{gather*}
  \hspace{-2em}\alpha_- \propto \sum_{i\in I}\sum_{g\in R} q(S_{i,g}=-1), \
  \alpha_0 \propto \sum_{i\in I}\sum_{g\in R} q(S_{i,g}=0), \
  \alpha_+ \propto \sum_{i\in I}\sum_{g\in R} q(S_{i,g}=+1), \\[2ex]
  \epsilon\propto  \sum_{i\in  I}\sum_{g\in T}  q(D_{i,g}=1),  \qquad
  (1-\epsilon)\propto \sum_{i\in I}\sum_{g\in T} q(D_{i,g}=0), \\[2ex]
  \mu_s=\frac{\sum_i\sum_g     q(S_{i,g}=s)    X_{i,g}}{\sum_i\sum_g
    q(S_{i,g}=s)}, \qquad
  \sigma_s^2=\frac{\sum_i\sum_g q(S_{i,g}=s)(\mu_s-X_i)^2}{\sum_i\sum_g q(S_{i,g}=s)}
\end{gather*}

\subsection*{Complexity analysis}
Step  M  only  involves  computing  a few  sums  of  size  [number  of
genes]$\times$[number of  samples] and is not time-consuming.
 Step E performs for each  sample a fixed number of passes
of Belief Propagation  in the factor graph. Each pass  consists in updating every  node with information
from  its  neighbors.   The  complexity of  updating  a  factor  grows
exponentially with its degree, therefore  it is important to limit the
number of variables of each factor.  It is done by replacing
the factors corresponding to the types $(4)$ and $(5)$ in Figure~\ref{factorgraph}   by tree-like structures with many
factors having 3 variables each.

With this approach the graph has approximately $N=2E+G$ nodes, where $E$ is the number of regulator-target edges in the regulation network, and $G$ the number of genes. A personal computer performs a few million node updates per second, thus step E will run in $t$ seconds if $N\times$[number of passes]$\times$[number of samples] is not much greater than $t$ millions.

\subsection*{Regulatory network inference from expression data}

To apply our methodology to real data, we use two different inference methods.

\paragraph*{LICORN.} The first one, named hLICORN, corresponds to the
LICORN model and is available in the CoRegNet Bioconductor
package~\cite{NRE15}. In a first step, it efficiently searches the discretized
gene expression matrix for sets of co-activators and co-repressors by frequent
items search techniques and locally selects combinations of co-repressors and
co-activators as candidate subnetworks. In a second step, it determines for
each gene the best sets among those candidates by running a regression. hLICORN
was shown to be suitable for cooperative regulation detection
\cite{ENB07,NRE15}.

\newcommand{\bbeta}{\boldsymbol\beta}
\newcommand{\hatbbeta}{\hat{\!\bbeta}}
\newcommand{\argmin}{\mathop{\mathrm{arg\ min}}}

\paragraph*{Cooperative-Lasso  +  Stability  Selection.}   The  second
inference procedure applies in a  continuous setup. It consists in two
steps: first, a selection step  performed with a sparse procedure; and
second, a resampling step whose  purpose is to stabilize the selection
for  more robustness  in  the reconstructed  network.   Here are  some
details.

\subparagraph*{Step  1: selection.}   For each  target gene,  a sparse
penalized  regression method  is used  to select  the set  of relevant
co-activators  and  co-inhibitors  among all  possible  transcription
factors. When  no special  structure is assumed  in the  network, this
task can be  performed with the Lasso penalty, as  it was successfully
applied for  network inference in  \cite{MB06}. Here, however,  we are
looking  for  sets  of  regulators that  work  group-wise,  either  as
co-activators or co-inhibitors.   To favor such a  structure, we build
on  the   penalty  proposed  in  \cite{CGA11,CGC12}   that  encourages
selection of predefined groups of variables sharing the same sign
(thus being either co-activators or co-inhibitors).  This regularization
scheme  is  known as  the  ``cooperative-Lasso''.   It was  originally
designed to work with  a set of groups that form  a partition over the
set of  regulators.  Here, we extend  this method to a  structure that
defines a hierarchy (or tree) on the set of regulators $R$ .
We denote  by $\mathcal{H}=\{\mathcal{H}_1,\dots,\mathcal{H}_K\}$ this
structure,  with $\mathcal{H}_k$  the  $k$th (non-empty)  node of  the
hierarchy.

Technically, the optimization problem  solved for selecting regulators
of gene $g$ is the following penalized regression problem
\begin{equation*}
  \label{eq:coop_linear}
  \hatbbeta^{(g)}            =           \argmin_{\bbeta^{(g)}\in\mathbb{R}^{|R|}}
  \frac{1}{2}\left\|\mathbf{X}_g-      \mathbf{X}_{R}      \bbeta^{(g)}
  \right\|^2 + \lambda
  \sum_{k=1}^K  \left\|\left(\bbeta^{(g)}_{\mathcal{H}_k}\right)^+\right\|_2 + \left\|\left(\bbeta^{(g)}_{\mathcal{H}_k}\right)^-\right\|_2,
\end{equation*}
with   $\mathbf{X}_g$  the   expression  profile   of  gene   $g$  and
$\mathbf{X}_R$ the expression  profiles of the regulators.
The parameter $\lambda>0$ tunes the amount of regularization, and thus
the number of regulators associated  with gene $g$; $\mathbf{v}^+$ and
$\mathbf{v}^-$ are the positive, respectively the negative elements of
a   vector   $\mathbf{v}$,    and   $\mathbf{v}_{\mathcal{H}_k}$   the
restriction of $\mathbf{v}$ to the elements in node $\mathcal{H}_k$ of
the hierarchy.  Hence, this  penalty favors selection of sign-coherent
groups of variables, like $(\bbeta_{\mathcal{H}_k}^{(g)})^+$, standing
for the estimated co-activators of gene $g$ in node $\mathcal{H}_k$ of
the    hierarchy,     or    $(\bbeta_{\mathcal{H}_k}^{(g)})^-$,    the
corresponding co-inhibitors.

\subparagraph*{Step2: Stabilization.}   We fit a sparse model as described above for each target gene, regressing on the same  set of regulators
$R$.  The  hierarchy $\mathcal{H}$ that we  used is obtained
by  performing  hierarchical  clustering  with average  linkage  on  a
distance based  upon the  correlation between expression  profiles. We
use the same $\lambda$ for each  gene, which is chosen large enough in
order to select  at least one set of regulators  for all target genes.
To select  the final edges  in the network,  we rely on  the stability
selection  procedure  of  \cite{meinshausen2010stability},  which  was
successfully  applied  to  the  reconstruction  of  robust  regulatory
networks in the case  of a simple Lasso penalty \cite{HMV12},  and is known
to  be   less  sensitive  than   selecting  one  $\lambda$   per  gene
(\emph{e.g}.   by  cross-validation).    This  technique  consists  in
refitting the regression model on  many subsamples obtained by drawing
randomly $n/2$ observations from the  original data set.  We replicate
10,000 times this operation and obtain a estimated probability of selection for
each edge. We fix  the threshold in order  to select a
number of  edges similar  to LICORN, which  corresponds to  edges with
a probability of selection greater than 0.65.

\section*{Results and Discussion}

\subsection*{Classification performances on simulated data sets}

In our experiments,  the score $q(D_{i,g}=1)$ is used  to determine if
gene  $g$  is deregulated  or  not  in  sample $i$.   Performances  are
evaluated with Precision-Recall (PR) curves, which are known to be more
informative than  ROC curves or accuracy  \cite{DG06} when considering
classification problem with very imbalanced data sets.

We generate expression data sets according to the model described earlier
 and feed them to the EM algorithm to evaluate its performance.
To study the  impact of each parameter, we try several
values  of this  parameter  while  all others  remain  fixed to  their
default  value.   Ten  data sets are generated and processed  in  each  setting,
resulting  in  $10$ PR  curves.   We  thus  obtain clouds  of  curves,
measuring both the  variability for a given parameter set and the
influence of the varying parameter.

 We unsurprisingly note that  $\sigma$ has
dramatic effect (see Figure~\ref{PRsigma}).  As a rule
  of thumb to distinguish two  states from one another, the associated
  standard  deviations must  be  smaller than  the difference  between
  their mean expressions.

 Meanwhile, large values of
$\epsilon$ mechanically result in better  PR: the more the deregulated
genes,  the  more  the  true positives  among  all  positives  (Figure
\ref{PReps}).

  On the contrary, all other parameter  have little effect on the performance
and we thus postpone the associated PR curves to
the Additional File 1. Those parameters are $\mu$,  $\alpha$,  the number of passes
in the Belief Propagation algorithm (as long as it is greater than five),  the number of genes and the
sample size (as long as their product is of several hundreds).




\subsection*{Managing the False Discovery Rate}

Consider  couples $(i,g)$ whose deregulation  score $q(D_{i,g}=1)=s$:
this  score  being  a  \emph{posterior}  probability,  the  expected
proportion of true (respectively false) positives is $s$ (respectively
$1-s$).   Similarly, if  $K$ pairs  pass the  threshold, the  expected
number  of true  positives  among them  is the  sum  of their  scores,
denoted  by $S$.   The  false discovery  rate  (FDR) may  be
estimated by $(K-S)/K$. In practice,  aiming for a particular FDR, one
can start with a threshold of 1  and lower it gradually: as more pairs
get selected,  the ratio  $(K-S)/K$ gradually
increases.  All  one has to  do is stop  when it reaches  the intended
FDR.   The  concordance   between  the intended  FDR  and the actual
proportion of false positives  is illustrated  on simulated  data sets
in  the Additional File 1.

\subsection*{Tests on real data}

We applied our method to the bladder cancer data set available in the
\texttt{R}-package  CoRegNet~\cite{NRE15}.    Expression  data  from
patients with different status was  pooled to infer gene co-regulatory
networks with  two independent  procedures, namely  \emph{hLICORN} and
the  hierarchical  \emph{Cooperative-Lasso}.   The  inferred  networks
reflect the  regulation trends  over the whole  set of  $184$ samples.
Our EM algorithm is then run using the same expression data, but since
samples are  now treated  individually, the  results reflect  how each
sample violates the regulatory rules generally followed by the others.

On  real   data,  the  true  deregulation   status  is
  unreachable. Hence, we match our  result with Copy Number Alteration
  (CNA) data collected from the same  samples, in order to support
  that our method correctly  identifies deregulated gene-sample pairs.
  We do  not expect CNAs  to precisely  coincide with failures  of the
  regulation network, so we do not  hope to detect exactly those pairs
  that present a  CNA. However, the number of  gene copies influences
the   expression   independently   from    expression   of   the   TFs
\cite{Pollack2002}.     We    therefore    expect   to    observe    a
link between CNA and gene deregulations.

To this end, we use CNA data provided by the CoRegNet package,
associating to each gene-sample pair a copy number state: $0$
  for the diploid state (two copies), $1$ for a copy number gain, $-1$
  for a copy number loss, and $2$ for a copy number amplification.
  Figure~\ref{boxplot} compares the distribution of the perturbation
  scores across copy number states by representing, for each copy number class, the
  empirical cumulative distribution function of the perturbation
  scores. For each value $s$ of the perturbation score in abscissa,
  the ordinate is the proportion of gene-sample pairs with a score
  greater than $s$. The fact that the curve corresponding to the
  diploid state is above all the other curves indicates that
  gene-sample pairs having a CNA are given a higher perturbation score
  diploid gene-sample pairs by our deregulation model.  Although the difference seems slight, it
  is highly significant given the large number of scores, as indicated
  by the $p$-value of the Student test for the pairwise differences
  between the diploid state and each of the other altered states. As
  expected, the scores of the ``amplification'' state $2$ are also higher
  than the scores of ``gain'' state $1$.


\section*{Conclusion}

In  the present  article,  we  develop a  statistical  model for  gene
expression based on a hidden  regulatory structure.  Given a reference
GRN, it allows to determine which  genes are misregulated in a sample,
meaning an  expression which  does not matches  the network  given the
expression  of its  regulators.   Numerical  experiments validate  the
algorithmic procedure: when applied to  bladder cancer data with known
CNA, the deregulation  score is higher in samples in  which genes have
an altered number of copies.

We believe  that our  methodology will be  useful to  understand which
regulation  mechanisms  are  altered  in  different  cancer  subtypes.
Indeed, the results of  our methodology are sample-specific.  However,
characterizing  the deregulations  which  are common  to  most of  the
individuals  suffering   a  given   cancer  subtype  is   a  promising
perspective.

The integration of  CNA to the methodology, as  already done in the
context   of  differential   expression~\cite{STP10},  will   also  be
considered  in future  work,  as it  would allow  a  better power  for
detecting genes suffering misregulation due to a copy alteration.


\begin{backmatter}

\section*{Availability of supporting data}

The EM algorithm described in this article is available as a Java archive at http://www.math-info.univ-paris5.fr/$\sim$ebirmele/

Bladder cancer data and hLicorn are available through the CoRegNet Bioconductor
package.

\section*{Abbreviations}

CNA: Copy Number Alteration
GRN: Gene Regulatory Network
PR curve: Precision-Recall
ROC curve: Receiver Operating Characteristic curve
TF: Transcription factor

\section*{Competing interests}
  The authors declare that they have no competing interests.

\section*{Author's contributions}
The work presented here was carried out in collaboration between all authors. ME and EB conceived the study. TP and EB designed it and wrote the manuscript. JC, PN and RN brought their expertise on inference and statistical interpretation on the real data. All authors provided valuable advises in developing the proposed method and modifying the manuscript. All authors read and approved the final manuscript.

\section*{Acknowledgements}
  The authors would like to thank Fran\c{c}ois Radvanyi for helpful discussions.

\section*{Declarations}
   This work  was partially supported  the CNRS (CREPE, PEPS BMI).
  Publication charges were funded by CHIST-ERA grant (AdaLab, ANR 14-CHR2-0001-01).


\bibliographystyle{bmc-mathphys} 
\bibliography{hal_article}      




\section*{Figures}
  \begin{figure}[h!]
  \includegraphics[width=.8\textwidth]{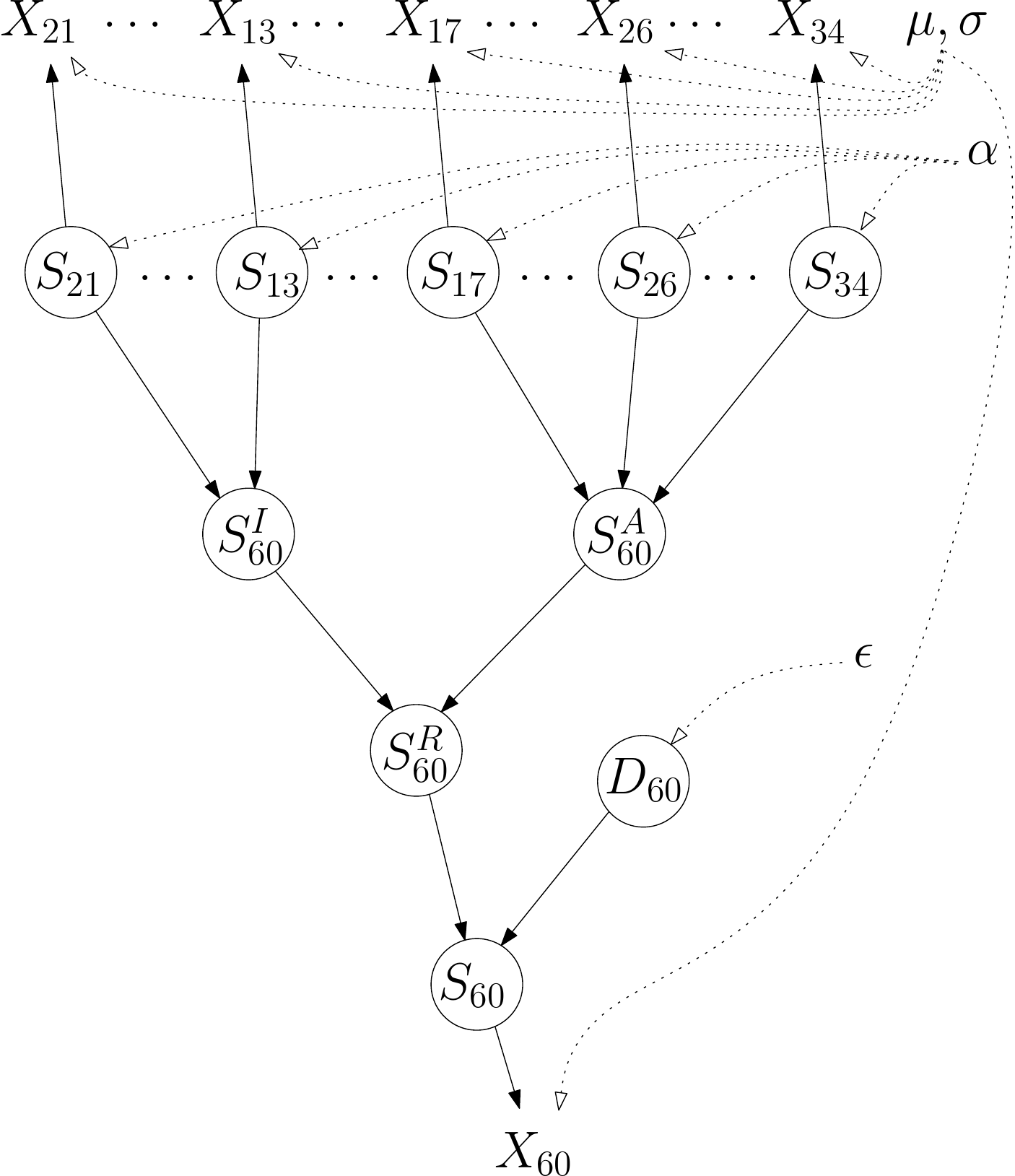}
  \caption{\csentence{The model for one target gene regulated by two co-inhibitors and three co-activators.}
       The circled variables are hidden. A dashed edge indicates that the distribution of the variable depends on the corresponding parameter.}
     \label{modelfig}
      \end{figure}

\begin{figure}[h!]
  \includegraphics[width=.8\textwidth]{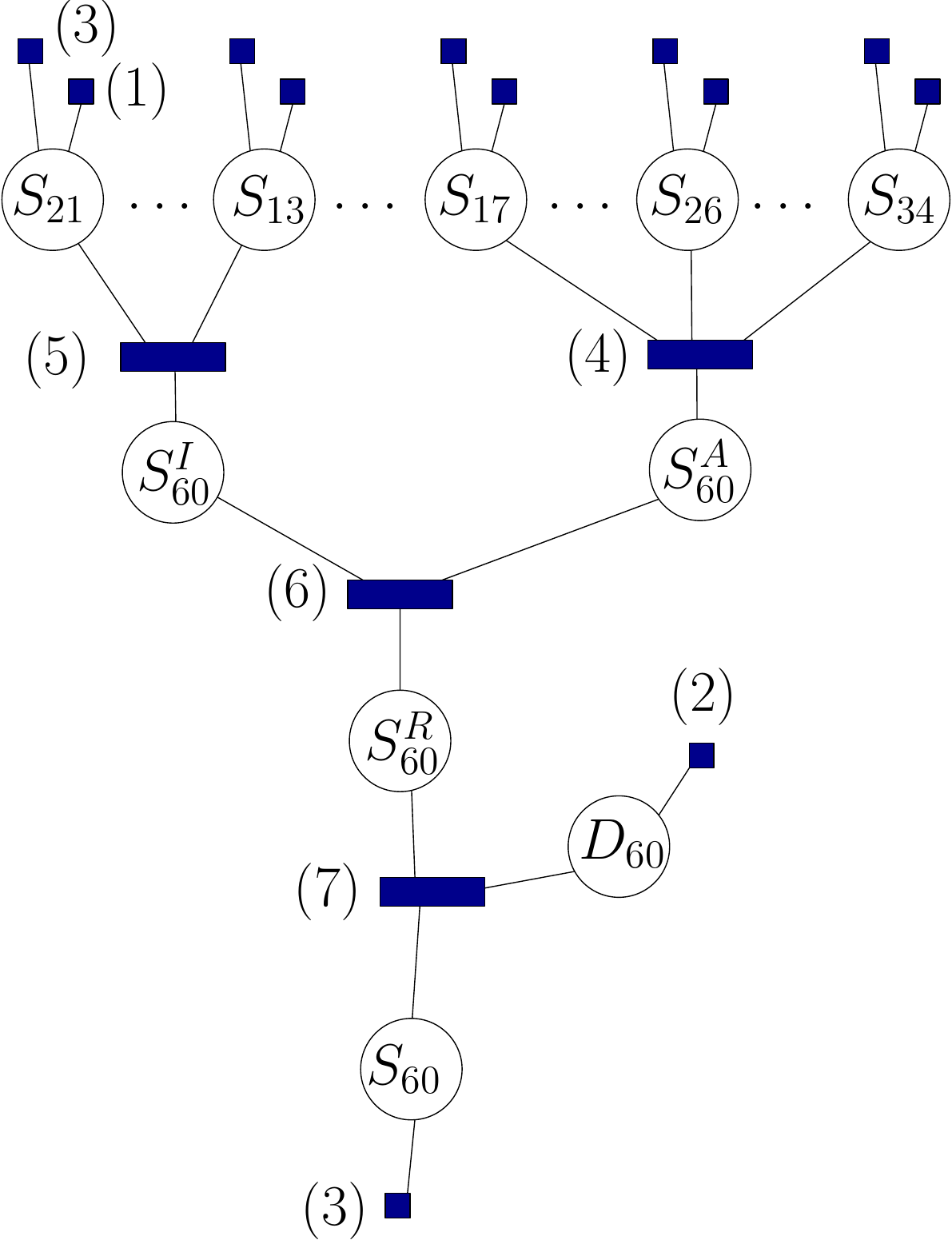}
  \caption{\csentence{A partial view of the factor graph.}
      The factor graph corresponding to Figure~\ref{modelfig}. The squares correspond to the factors, and are numbered according to the text. The algorithm oteratively updates the distribution of the circled variables.}
    \label{factorgraph}
  \end{figure}

\begin{figure}[h!]
  \includegraphics[width=.8\textwidth]{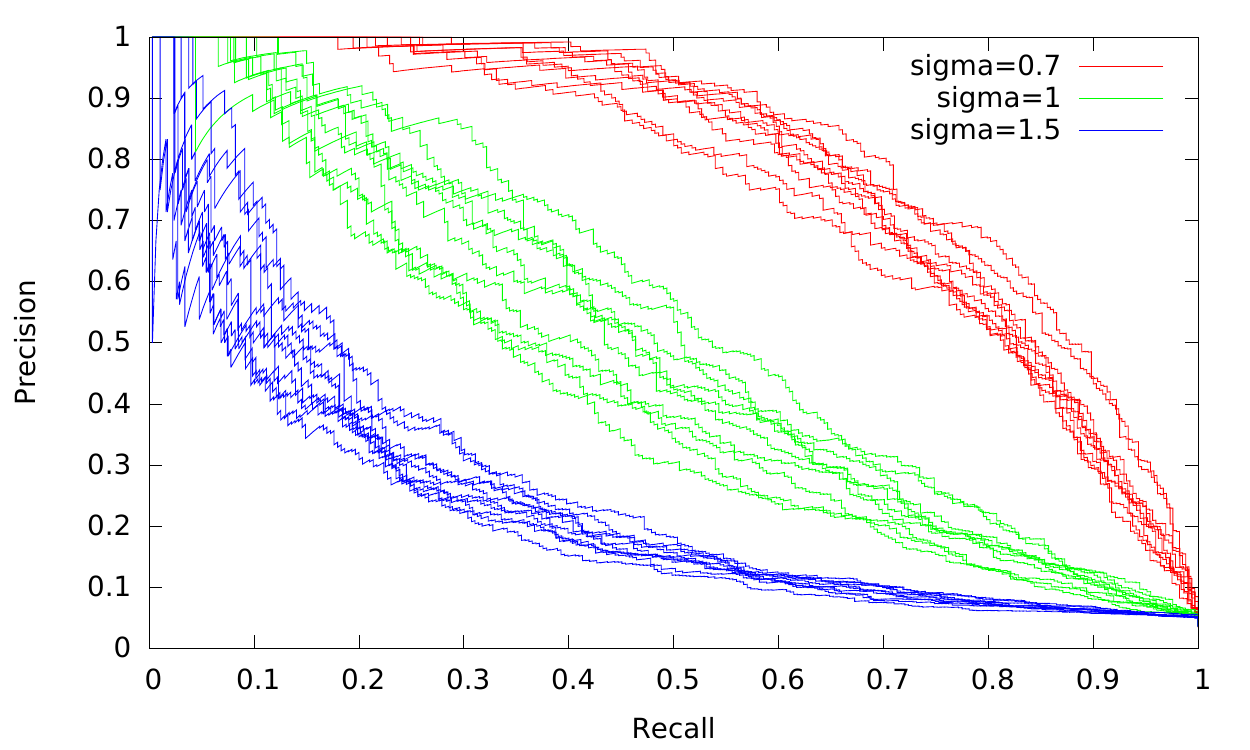}
  \caption{\csentence{Influence of $\sigma$.}
      PR curves for simulations with varying $\sigma$, with means $(\mu_-,\mu_0,\mu_+)=(-1,0,1)$. Ten simulations are run for each value}
    \label{PRsigma}
  \end{figure}

\begin{figure}[h!]
  \includegraphics[width=.8\textwidth]{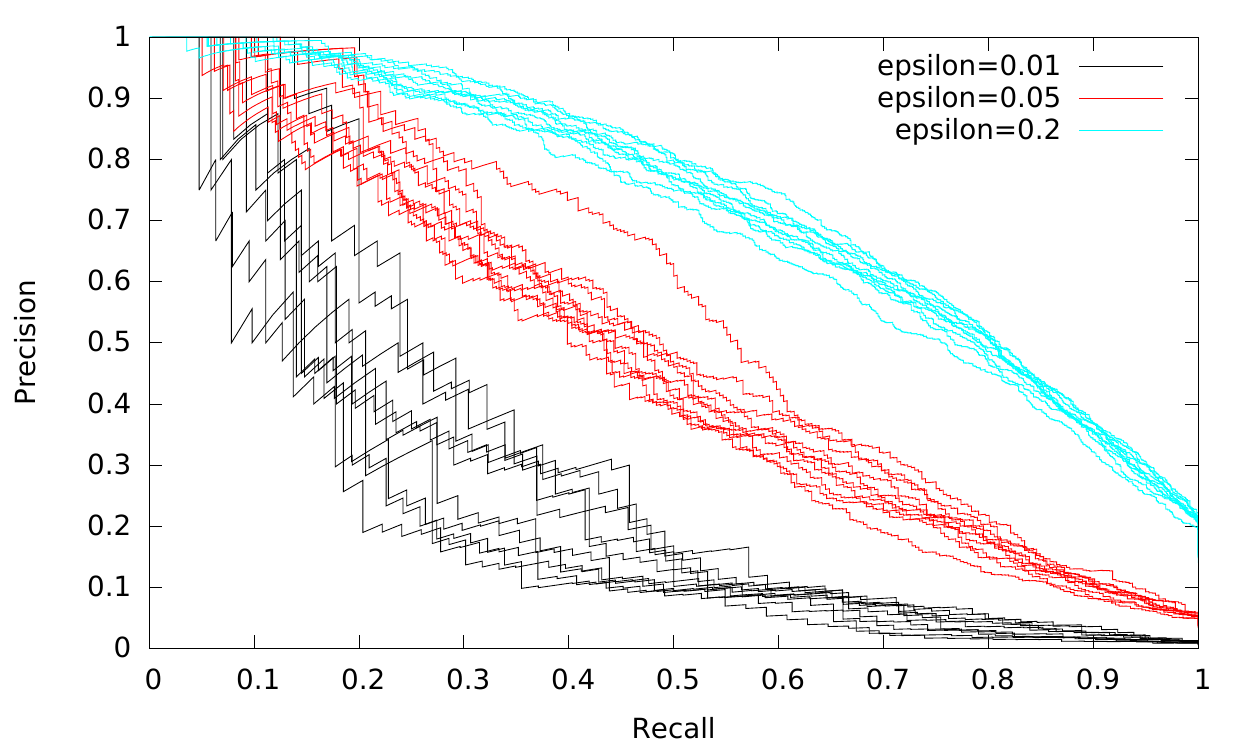}
  \caption{\csentence{Influence of $\epsilon$.}
      PR curves for simulations with varying $\epsilon$. Ten simulations are run for each value}
    \label{PReps}
  \end{figure}

\begin{figure}[h!]
  \includegraphics[width=.8\textwidth]{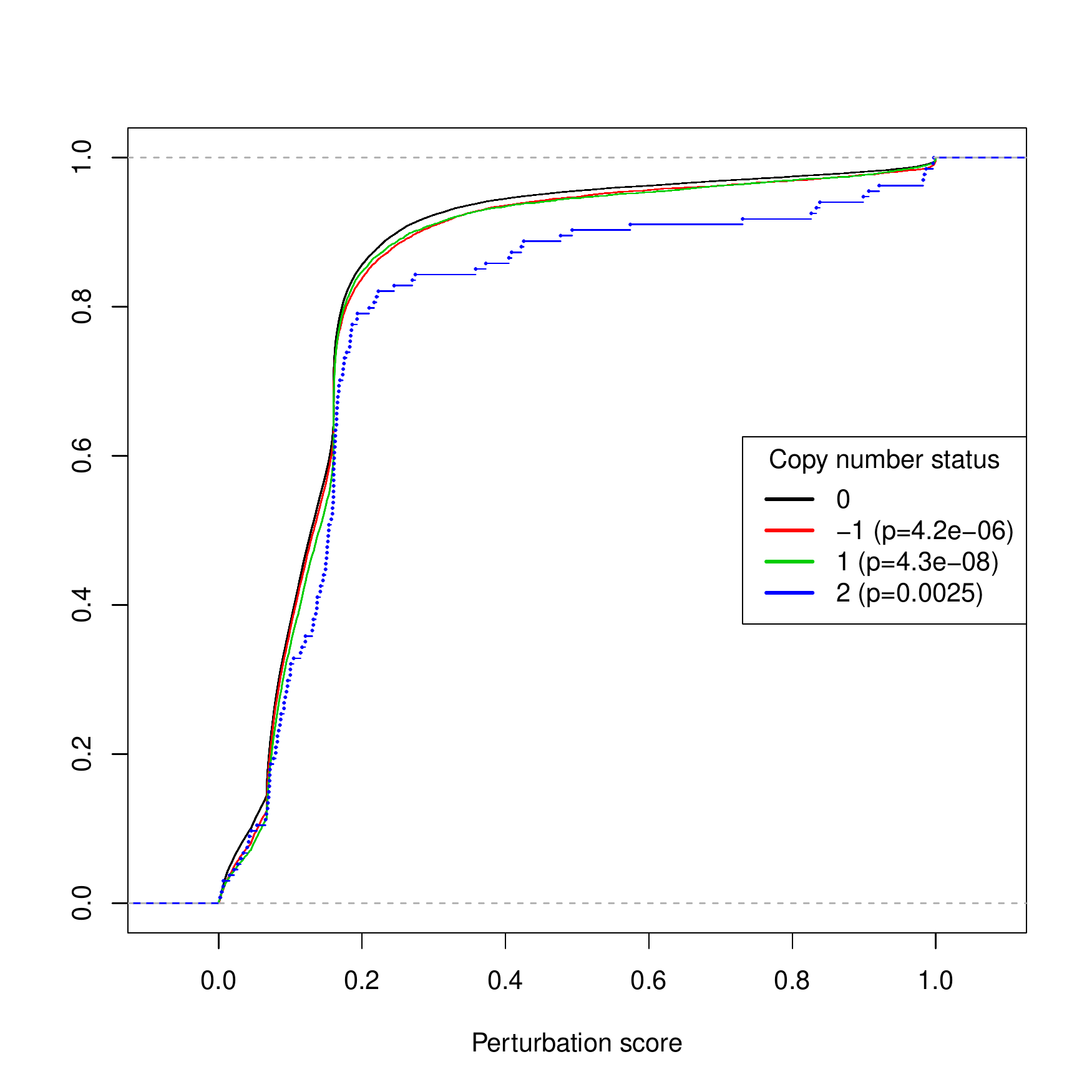}
  \caption{\csentence{Empirical cumulative distribution of scores, by Copy-Number status.}
    Student's test is used to compare every altered state with the normal.}
  \label{boxplot}
  \end{figure}


\section*{Tables}
\begin{table}[h!]
\caption{LICORN truth table. How the target gene behaves (unless it is deregulated) according to its co-activators' state $A$ and co-inhibitors' state $I$. }
 \begin{tabular}{c|c|c|c|}
  \diaghead(3,-2){\hspace{2em}}{I}{A} & - & 0 & + \\ 
  \hline
  - & 0 & + & +\\
  \hline
  0 & - & 0 & +\\
  \hline
  + & - & - & -\\
  \hline
 \end{tabular}
 \label{licorntable}
\end{table}


\section*{Additional Files}
  \subsection*{Additional\_File\_1.pdf}
    File containing PR curves for varying $\alpha$, $\mu$, the number of genes/samples and the number of belief propagation iterations. It also contains figures illustrating the FDR estimation on simulated data.

\end{backmatter}
\end{document}